# Magnon-magnon interaction induced by dynamic coupling in a hybrid magnonic crystal

Rawnak Sultana[1], Mojtaba Taghipour Kaffash[1], Gianluca Gubbiotti[2], Yi Ji[1], M. Benjamin Jungfleisch[1], and Federico Montoncello[3]

[1]*Department of Physics and Astronomy, University of Delaware, Newark, DE19716, USA*

[2] *CNR-Istituto Officina dei Materiali (IOM), Unità di Perugia, 06123 Perugia, Italy*

[3]*Dipartimento di Fisica e Scienze della Terra, Università di Ferrara, 44121 Ferrara, Italy*

**ABSTRACT** We report a combined experimental and numerical investigation of spin-wave dynamics in a hybrid magnonic crystal consisting of a CoFeB artificial spin ice (ASI) of stadium-shaped nanoelements patterned atop a continuous NiFe film, separated by a 5 nm $Al_2O_3$ spacer. Using Brillouin light scattering spectroscopy, we probe the frequency dependence of thermal spin waves as functions of applied magnetic field and wavevector, revealing the decisive role of interlayer dipolar coupling in the magnetization dynamics. Micromagnetic simulations complement the experiments, showing a strong interplay between ASI edge modes and backward volume modes in the NiFe film. The contrast in saturation magnetization between CoFeB and NiFe enhances this coupling, leading to a pronounced hybridization manifested as a triplet of peaks in the BLS spectra—predicted by simulations and observed experimentally. This magnon-magnon coupling persists over a wide magnetic field range, shaping both the spin-wave dispersion at fixed fields and the full frequency-field response throughout the magnetic hysteresis loop. Our findings establish how ASI geometry can selectively enhance specific spin-wave wavelengths in the underlying film, thereby boosting their amplitude and identifying them as preferential channels for magnonic signal transport and manipulation.

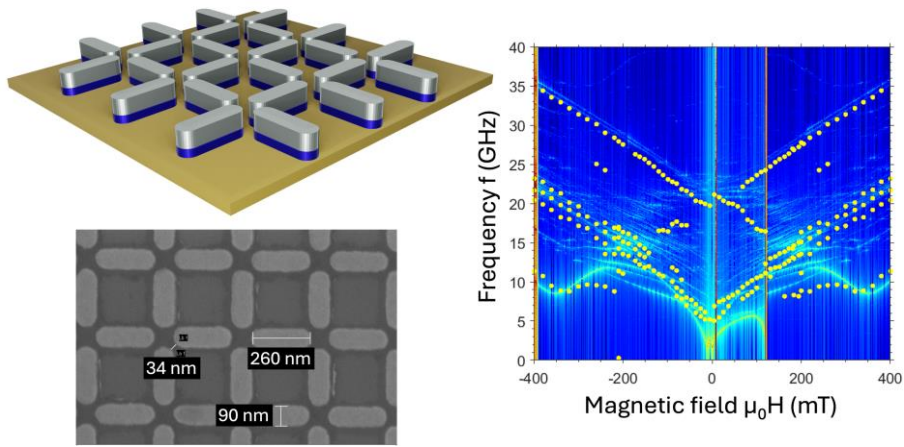

## INTRODUCTION

Artificial spin ice (ASI) structures consist of arrays of nanoscale magnetic elements arranged in geometrical patterns that mimic the frustration observed in natural spin ice.[1,2,3] They represent a fascinating class of engineered metamaterials, specifically designed to control the spin-wave (SW) band structure, paving the way for the development of magnonic devices with reconfigurable magnetic properties, enabling tunable properties on demand.[4,5,6,7,8,9] A critical aspect of ASI behavior is the strong magnon-magnon coupling effects,[10,11] which arise due to strong dipolar coupling between adjacent nanomagnetic elements.[12,13]

In previous works, we studied the effect of NiFe ASI deposited on top of an unpatterned NiFe film with varying spacer layer thickness, composed of aluminum oxide ($Al_2O_3$),[13] to understand how the dynamic stray magnetic fields generated by the ASI affect the SW propagation in an underlying NiFe thin film.[14,15] In those works, ASI and film were made of the same material, hence having the same saturation magnetization ($M_S$). This was found to facilitate the dynamic coupling between the film backward volume SW and several ASI bulk modes, leading to robust ASI–film hybrid modes that persist across variations in both field and thickness.

Here, we present a novel variation of the ASI/film hybrid structure, employing different constituent materials to uncover new coupling phenomena. In the present work, the ASI is made of high saturation magnetization ($M_s$) CoFeB, while the underlying thin film consists of soft NiFe, characterized by its comparatively lower $M_s$. We reveal evidence for a magnon-magnon interaction between SWs in the film and the ASI edge modes, an interaction typically considered negligible due to weak dynamic stray fields associated with localized modes. The pronounced difference in the two material parameters displaces the ASI bulk mode frequencies relative to the film spin waves, effectively preventing their coupling. However, the ASI edge modes remain within the same frequency range as the film SWs, enabling a distinct dynamic coupling. As a result, the otherwise negligible magnon-magnon interaction between edge modes and film SW is enhanced, thereby broadening the range of potential applications in reconfigurable magnonic devices and computing systems.

We use Brillouin light scattering (BLS) spectroscopy to measure the spectra of thermally excited SWs as a function of the magnetic field strength and wavevector. Furthermore, we employ micromagnetic simulations to interpret the experimental findings. This two-pronged approach allows us to reveal how the $M_S$ difference between the ASI and the film can be effectively utilized to modulate the strength of the magnon-magnon interaction. In particular, since the saturation magnetization of a single layer can be tuned—for example, by laser irradiation[16,17,18]—it becomes possible to engineer a magnonic device in which, through suitable material choice and design, the film mode can be dynamically coupled either to ASI edge modes or to bulk modes, depending on the irradiation.

Our work uncovers a new degree of freedom in designing tunable and energy-efficient magnonic systems, contributing to the emerging research field of 3D magnonics,[19] where vertical stacking offers versatile coupling conditions for controlling SW propagation in patterned[20,21] and unpatterned ferromagnetic films.[22,23,24]

**RESULTS AND DISCUSSION**

The fabricated samples consist of the following layers:

#1 A continuous NiFe film with a thickness of 20 nm.

#2 A continuous CoFeB film with a thickness of 20 nm.

#3 A square single-layer ASI lattice composed of 20-nm-thick CoFeB stadium-shaped nanoelements (islands).

#4 A hybrid structure consisting of a square ASI lattice (composed of 20-nm-thick CoFeB islands) separated from a 20-nm-thick NiFe film by a 5-nm-thick $Al_2O_3$ non-magnetic spacer layer.

The CoFeB ASI islands are designed as stadium-shaped elements with nominal lateral dimensions of $260 \times 90$ nm². Figure 1 (a) shows a graphical representation of the sample structure with emphasis on the vertical materials stack while in (b) a representative scanning electron microscopy (SEM) image of the ASI sample is shown. These islands are arranged in a square lattice, maintaining a minimum edge-to-edge spacing of 34 nm. The nominal lattice constant, defined as the center-to-center distance between neighboring nanomagnets belonging to the same sublattice, is $a$=350 nm. This spacing determines the Brillouin zone (BZ) boundary at a wavevector $k_{BZ} = \pi/a \approx 0.9 \times 10^7$ rad/m.

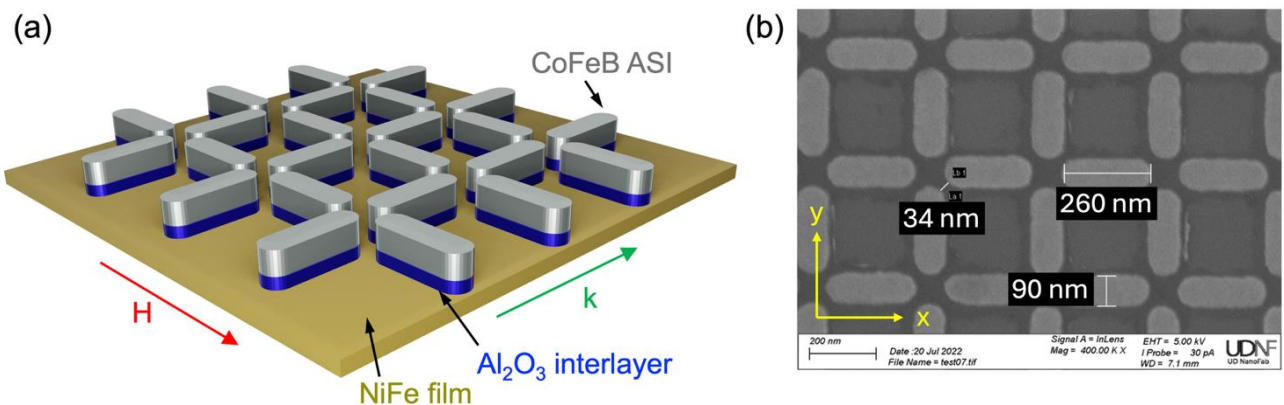

**Figure 1. (a) Graphical representation of sample structure consisting of CoFeB ASI deposited on top of a NiFe film separated by a 5 nm thick $Al_2O_3$ spacer. (b) Representative SEM image of the ASI sample along with coordinate axis and the lateral dimensions of the ASI islands.**

The measured MOKE loops are presented in Fig. 2. For the unpatterned NiFe film [Sample #1, Fig. 2(a)], the measured MOKE loop exhibits a step-function-like shape, where the magnetization remains nearly saturated over the entire field range, except for a narrow (~1 mT) interval around $\mu_0 H=0$, where the magnetization abruptly switches to negative values. The loop for the CoFeB film [Sample #2, Fig. 2(b)] displays a similar shape, but a larger coercivity of approximately 15 mT.

For the isolated CoFeB ASI [Sample #3, Fig. 2(c)], starting from negative saturation ($\mu_0 H=-250$ mT), the MOKE loop exhibits an elongated shape, characterized by a gradual decrease in magnetization as the field magnitude decreases. This behavior suggests a progressive re-alignment of the magnetic moments in the horizontal islands toward the easy axis (along the $x$-direction) without a clear indication of distinct switching fields for the families of CoFeB islands either parallel or perpendicular to the applied magnetic field.

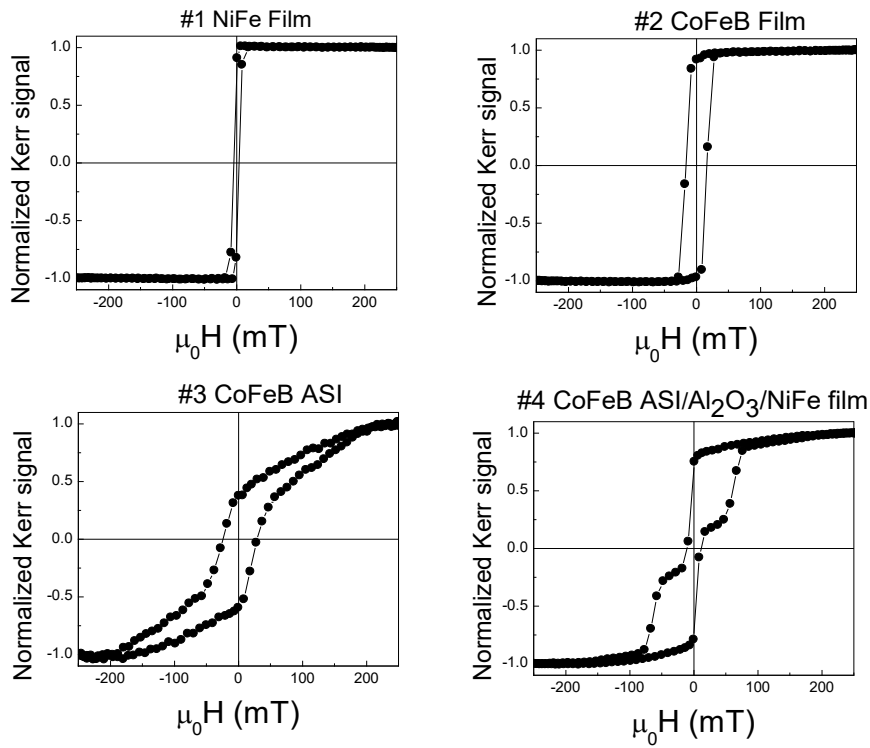

**Figure 2. Longitudinal MOKE loops for (a) the continuous NiFe film (Sample #1), (b) the continuous CoFeB film (Sample #2), (c) the ASI array of CoFeB islands (Sample #3) and (d) the hybrid CoFeB ASI/$Al_2O_3$/NiFe film structure (Sample #4).**

For the hybrid system [Sample #4, Fig. 2(d)], starting from $\mu_0H=-250$ mT, the MOKE loop initially exhibits a gradual increase in magnetization as the field magnitude is ramped up to zero field. This behavior suggests a progressive alignment of the magnetic moments in the vertical islands toward the field direction (*x*-axis), similar to what is observed for the isolated ASI [Sample #3, Fig. 2(c)]. As the field is swept across $\mu_0H=0$ mT, a sharp jump in magnetization occurs, which is attributed to the reversal of the continuous NiFe film. For positive fields, the loop features an almost flat plateau, extending from +18 to about 50 mT, followed by a second sharp magnetization switching event associated with the reversal of the vertical CoFeB ASI islands, which are oriented along the direction of the externally applied field. Above 80 mT, a gradual approach to saturation, reached at about 250 mT, is observed. It is noteworthy that, despite the finite penetration depth of light, the MOKE signal is clearly detected from the uncovered regions of the NiFe film that are not overlaid by the ASI structures. A similar shape of the MOKE hysteresis loop was previously observed in the case of a NiFe ASI patterned on top of a continuous NiFe film.[15]

BLS spectra were acquired by sweeping the magnetic field $\mu_0H$ in the range from -400 to +400 mT at normal light incidence (determining a SW wavevector $k=0$) and then by varying the wavevector $k$ from 0 to $2.0\times10^7$ rad/m at fixed $\mu_0H = 350$ mT. Representative BLS spectra recorded at $k=0$ (normal light incidence) and $\mu_0H=350$ mT for all the investigated samples are shown in Fig. 3.

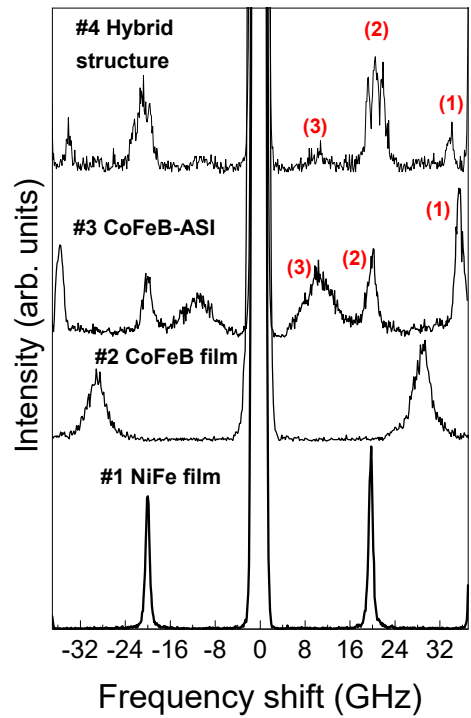

**Figure 3, BLS spectra (intensity vs frequency shift) for all samples under investigation measured at normal light incidence (k=0) and at a fixed magnetic field of $\mu_0 H$=350 mT. Spectra are vertically shifted to highlight the different peak frequencies in the investigated samples. For samples #3 and #4, peaks are labeled by integer numbers 1, 2, and 3 to follow their field and $k$ evolution and discussion in the text.**

For the unpatterned NiFe and CoFeB thin films (Samples #1 and #2, respectively), spectra exhibit a single prominent peak corresponding to the Damon-Eshbach (DE) mode. The frequency of these peaks is higher for CoFeB than for NiFe, primarily due to CoFeB's larger saturation magnetization. Furthermore, the CoFeB film shows a significantly broader peak compared to NiFe, indicating increased damping or inhomogeneous linewidth broadening of the CoFeB sample.

In contrast, the CoFeB ASI structure (Sample #3) exhibits three distinct peaks, labeled 1, 2, and 3. The two high-frequency peaks (1 and 2), located at approximately 35.5 GHz and 20.2 GHz, are sharp and well-defined, whereas the low-frequency peak (3) is considerably broader, extending from about 7.5 to 13.5 GHz, indicating contributions from edge modes whose frequencies are sensitive to the island geometry, including shape, size, and edge roughness. Notably, peak (2) for the ASI sample (20.00 GHz) is very close in frequency to the peak observed in the NiFe film (19.65 GHz). This small frequency difference has important implications, which will be presented and discussed in the following sections.

In the hybrid sample #4, three peaks are also observed. The low-frequency peak at 10.2 GHz is notably broad and low in intensity. Upon closer examination, peak 2, located near 20 GHz, splits into a triplet of closely spaced peaks, indicating a more complex resonance structure. This splitting suggests the presence of additional mode interactions, likely resulting from the coupling between the localized modes of the CoFeB ASI islands and the propagating SWs in the NiFe continuous film, where the resonances of the film and the ASI are nearly degenerate. This observation will be discussed in greater detail below.

The NiFe and CoFeB magnetic parameters were obtained by fitting the experimental frequency/wavevector dependence of the unpatterned films with the dipole-exchange surface SW dispersion relation:[25]

$$\omega^2 = \omega_{ex}(\omega_{ex} + \omega_M) + \frac{\omega_M^2}{4}[1 - e^{-2kL}] \qquad \text{Eq. (1)}$$

where $k$ is the SW wavevector perpendicular to the applied magnetic field $H$, i.e., the so called DE[26] configuration, $\omega_{ex} = \gamma\mu_0 H + \lambda_{ex}^2 \omega_M k^2$, with $\omega_M = \gamma\mu_0 M_s$ and $\lambda_{ex}^2 = \frac{2A}{\mu_0 M_s^2}$ (square of the exchange length), and $L$ is the thickness of the continuous reference medium, $M_s$ is the saturation magnetization, $A$ is the exchange stiffness parameter, and $\omega = 2\pi\nu$, where $\nu$ is the SW frequency.

From the best fit [Fig. 4(a)], we obtain $M_s$= 800 kA/m, $A$ =13 pJ/m for NiFe, and $M_s$=1620 kA/m, $A$ =75 pJ/m for CoFeB. The gyromagnetic ratio was set to $\gamma$ = 185 rad GHz/T for both materials. We cross-checked the thicknesses of the two thin film layers by treating them as fitting parameters. This resulted in a value of 15 nm, slightly lower than the nominal thickness values measured using a thickness monitor during deposition (likely due to a thin magnetic dead layer).

With the above parameters, the frequency/field curves simultaneously fit the experimental measurements [Fig. 4(b)], showing a monotonic dependence on the applied field with a symmetric behavior for opposite field values, characteristic of the DE SW in single-layer magnetic films. As expected, the measured resonance frequency of the CoFeB film is consistently higher than that of the NiFe film due to the larger saturation magnetization of CoFeB.

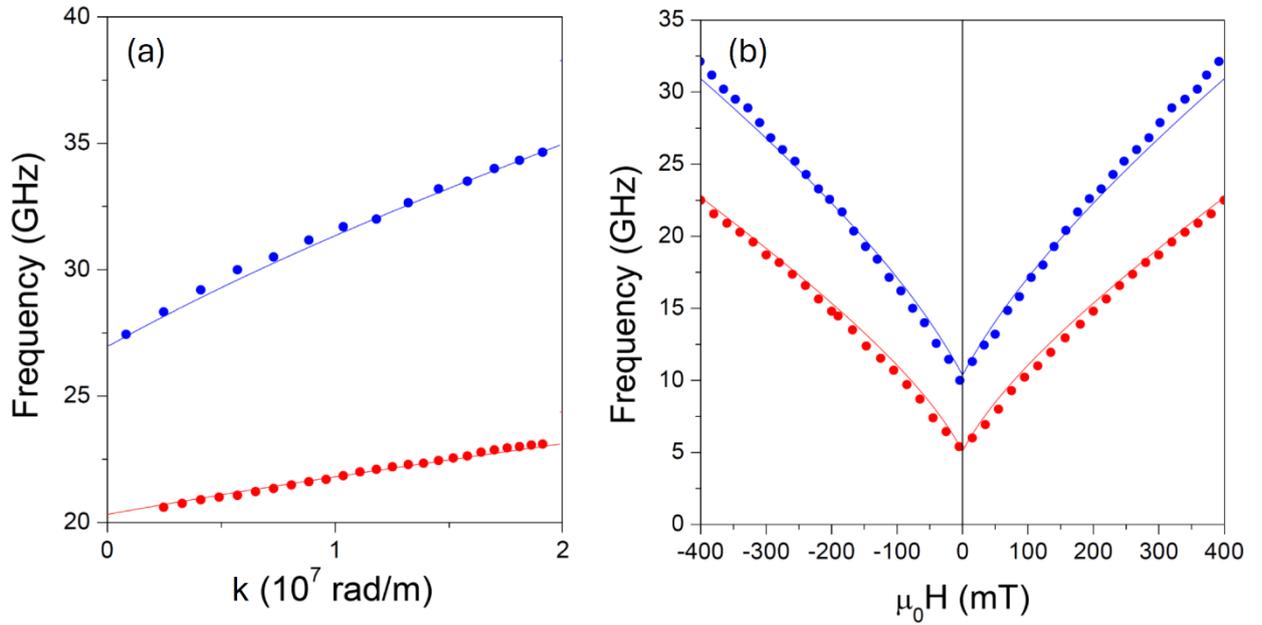

**Figure 4. Measured (symbols) and analytical (lines) dispersion curves taken at $\mu_0 H$=350 mT (a), and frequency/field curves taken at an incidence angle of 10° (i.e $k$=0.41×10$^7$ rad/m) (b) for the NiFe (red) and CoFeB (blue) reference films. The analytical fits to the experimental data are used to extract the magnetic parameters employed in the micromagnetic simulations.**

We begin by discussing the frequency-field dependence, comparing the experimental measurements to the simulated results (Fig. 5). In the experimental BLS spectra, we tracked the field-dependent evolution of the magnonic peaks, shown at 350 mT in Fig. 3, as a function of the applied field. For this purpose, we first saturated the sample magnetization by applying a strong external magnetic field of -400 mT along the *x*-direction. We then systematically ramped the field up to +400 mT in steps of 20 mT, following the ascending branch of the MOKE hysteresis loop shown in Fig. 2.

For the single-layer ASI sample (Sample #3), three peaks are observed in the BLS spectra at negative saturation as shown in Fig. 5 (a). Comparison with the simulations, based on the calculated mode frequencies and the spatial profiles of modes shown in Fig. 6, suggests that these peaks correspond to the following modes (from highest to lowest frequency): the fundamental mode of the islands aligned parallel to the applied field [$F_\parallel$ in Fig. 6(f)], the edge mode with zero nodes of the island parallel to the field [0-$EM_\parallel$ in Fig. 6 (d)], and the edge mode without nodes of the island perpendicular to the applied field [0-$EM_\perp$ in Fig. 6(b)]. The latter mode is particularly intense and extend into the central part of the island, likely due to hybridization with the fundamental mode of the same island. Interestingly, the fundamental mode of the island perpendicular to the applied field ($F_\perp$) does not appear as distinct, stand-alone mode at large negative fields. Instead, due to this hybridization, mode 0-$EM_\perp$ effectively also serve as the $F_\perp$ mode: this feature holds for a wide range of field values, until around -160 mT, when an independent $F_\perp$ arises due to magnetization rotation (see below).

Since peaks (1) and (2) in Fig. 3 originate from the islands parallel to the applied field, in which the magnetization distribution is almost unchanged up to -200 mT, they follow the typical linear Larmor frequency behavior $\omega = \gamma \mu_0 H$ (in the simulations down to $\mu_0 H=0$). Conversely, peak (3) stems from the island perpendicular to the applied field, whose magnetization undergoes a gradual reorientation (due to shape anisotropy), and hence it displays the typical "W-shape" behavior described in Refs.[27,28]: its curve intersects with peak (2) twice, at $\mu_0 H$ = -160 mT and $\mu_0 H$ = 0, as indicated by the red circle in Fig. 5(a).

In the simulations, a remarkable behavior is observed for the fully antisymmetric edge mode (1x1)-EM [i.e., an edge mode, with zero intensity in the bulk, having one node along either axis, Fig. 6(g)]. Due to the gradual rotation of the underlying magnetization and the formation of an S-state configuration,[27,29] this mode also undergoes a progressive transformation acquiring a more symmetric profile. As the field approaches -160 mT (i.e., as a magnetization rotation of 45 degrees is approached), the mode intensity in the bulk increases and, eventually, this edge mode transforms into a bulk mode, namely the new $F_\perp$ (correspondingly, the 0-$EM_\perp$ gradually loses intensity in the bulk and hence the role of fundamental mode). This effect appears to be confirmed by the experiments [Fig. 5(a)], where a BLS signal emerges at around 19 GHz at -100 mT, in excellent agreement with the simulated frequency of the $F_\perp$ mode. The evolution of this transformation is illustrated by the simulated mode profiles at -350 mT, -250 mT and -100 mT in Fig. 6 (g,h,i).

In the simulations, 0-EM$_\perp$ and 0-EM$_\parallel$ intersect with one another at two field values: $\mu_0 H = -160$ mT (in Fig. 5(a), the leftmost red circle) and $\mu_0 H = 0$ (in Fig. 5(a), the lower red circle). At both fields, the magnetizations of the islands oriented parallel and perpendicular to the applied field are identical and, thus, indistinguishable. Specifically, at $\mu_0 H = -160$ mT, the magnetization in all islands (regardless of their orientation) adopts an S-state configuration with a 45° angle relative to the applied field, as previously discussed. At $\mu_0 H = 0$, the magnetization aligns along the long axis of each island, again making the two orientations equivalent. Similarly, F$_\perp$ and F$_\parallel$ become degenerate at $\mu_0 H = 0$ (in Fig. 5(a), the higher red circle), where the magnetization aligns along either island long axis, hence making them indistinguishable. Considering these degeneracy effects, the latter, involving the fundamental modes, is clearly detected in the experiment (Fig. 5(a), the higher red circle) at a frequency of about 20.5 GHz. In contrast, the former degeneracy, involving the edge modes, is less evident in the experiment [lower red circle in Fig. 5(a)]. In fact, if we inspect the BLS measurements only (symbols in Fig. 5), the first edge-mode intersection appears to occur at a lower field, i.e., around $\mu_0 H = -250$ mT (−160 mT in simulations), where the extrapolated trend of the 0-EM$_\perp$ points seems to intersect that of 0-EM$_\parallel$ (Fig. 5(a), dashed oblique line). If this were true, the experimental data would fail to reproduce the second EM degeneracy, expected at $\mu_0 H = 0$: the trend of the dashed oblique line (0-EM$_\parallel$) would intersect the frequency axis ($\mu_0 H = 0$) at approximately 3.5 GHz, appearing disconnected from the trend inferred from the experimental points measured within the interval [-150, -100] mT at around 15 GHz for the 0-EM$_\perp$ mode.

Despite being difficult to unambiguously interpret, we justify this discrepancy recalling that edge modes are typically very sensitive to the geometrical parameters (shape, edge roughness, spacing, etc.): therefore, because of fine mismatches between simulated and real samples, the measured frequency/field slope of mode 0-EM$_\parallel$ appears to significantly depart form the simulated trend and the actual BLS intensity of these modes appear below the noise level.

In the simulations, we also observe higher-order modes, which are close in frequency to F$_\parallel$. These modes exhibit $m$ nodes perpendicularly to the magnetization and are identified as "backward" modes ($m$-BA) akin to the backward-volume SW configuration.[15,30] The most intense among these are the modes with the lowest even values of $m$ [$m$=2 in Fig. 6(e)]. In the real sample, these higher-order modes may merge, resulting in a broad peak (1) observed in the experimental spectra in Figs. 3 and 5.

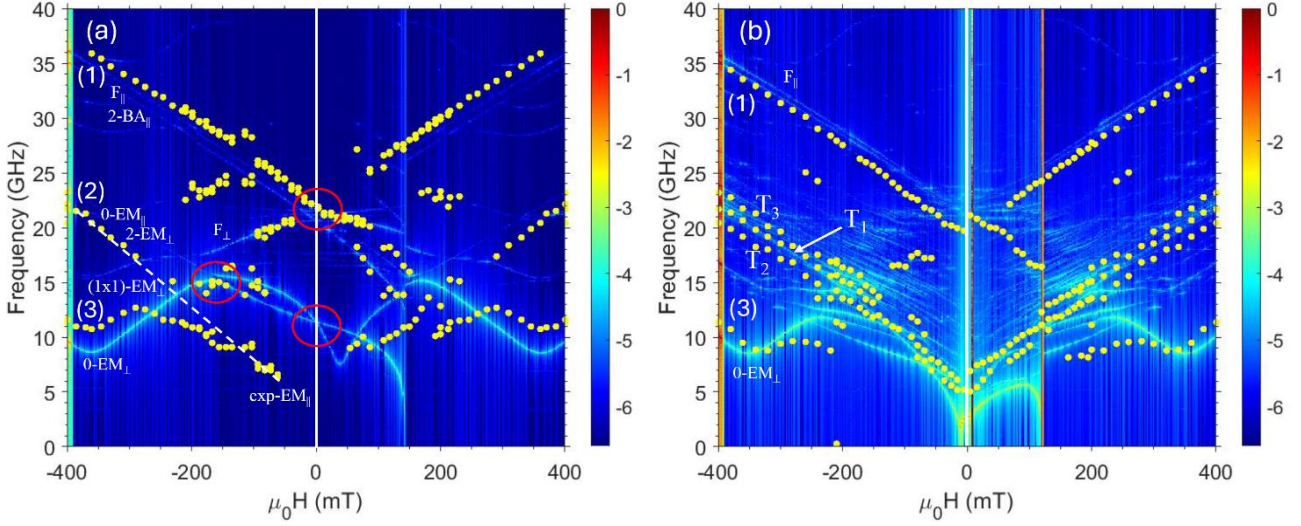

**Figure 5.** Field dependence (with increasing field from left to right) of the SW frequencies (full point) derived in the BLS spectra measured at *k*=0 rad/m for (a) the CoFeB ASI sample (Sample #3) and (b) the hybrid CoFeB ASI/Al$_2$O$_3$/NiFe film sample (Sample #4). The color maps (in arbitrary units) represent the square of the average Fourier coefficients of all layers in the 2x2 supercell. Panel (a): the oblique dashed line serves as a visual guide, indicating the expected trend of the experimental data for 0-EM$_\parallel$; the leftmost red circle at -160 mT marks the first point at which the edge modes of both islands are degenerate in frequency in the simulation: this results in an increased intensity, which might correspond to the concentration of experimental data points observed in the same region. The rightmost red circles highlight the crossing points where edge (lowest circle) and fundamental (highest circle) modes become degenerate for all the islands. Panel (b): hybrid modes resulting after ASI/Film magnon-magnon interaction are labeled T$_1$, T$_2$ and T$_3$, following the order in frequency and intensity found in the simulations, T$_1$ being the most intense one. However, the experimental results - based on the measured BLS intensities - indicate that T$_1$ and T$_2$ must be interchanged.

Finally, we observe that mode 2-EM$_\perp$, [Fig. 6(c)] is close in frequency (less than 1 GHz) and weakly hybridized with mode 0-EM$_\parallel$ [Fig. 6(d)]. This proximity will be significant in the following discussion, where we examine the coupling between ASI and film modes. Here, we want to recall how ASI modes generally arise as linear combinations (superpositions) of the normal modes of the individual (isolated) nanoelements (islands), which serve as the fundamental building blocks for the interacting ASI system[28]. The dipolar interaction among the ASI islands introduces a perturbation that couples the single-island wavefunctions and shifts their frequencies.

The overall frequency-field dependence of the ASI-film hybrid structure (Sample #4), as shown in Fig. 5(b), closely resembles that of the isolated ASI structure, particularly with respect to the peaks 1 and 3. We argue that these peaks are due to excitations confined to the ASI layer only, a point we will substantiate further below. However, an intriguing feature emerges for peak 2, where a triplet of peaks, labeled as $T_1$, $T_2$ and $T_3$, is observed over a large field range instead of the single peak measured for the ASI structure. The label order, $T_1$, $T_2$ and $T_3$, follows the order in frequency and intensity found in the simulations, $T_1$ being the most intense one (also dispersive, as will be presented below). However, the experimental results in Fig. 9(b) indicate - based on the measured BLS intensities - that $T_1$ and $T_2$ must be interchanged. This explains why $T_1$ appears in the middle of the triplet in Fig. 5(b), as it will also in Fig. 10(b).

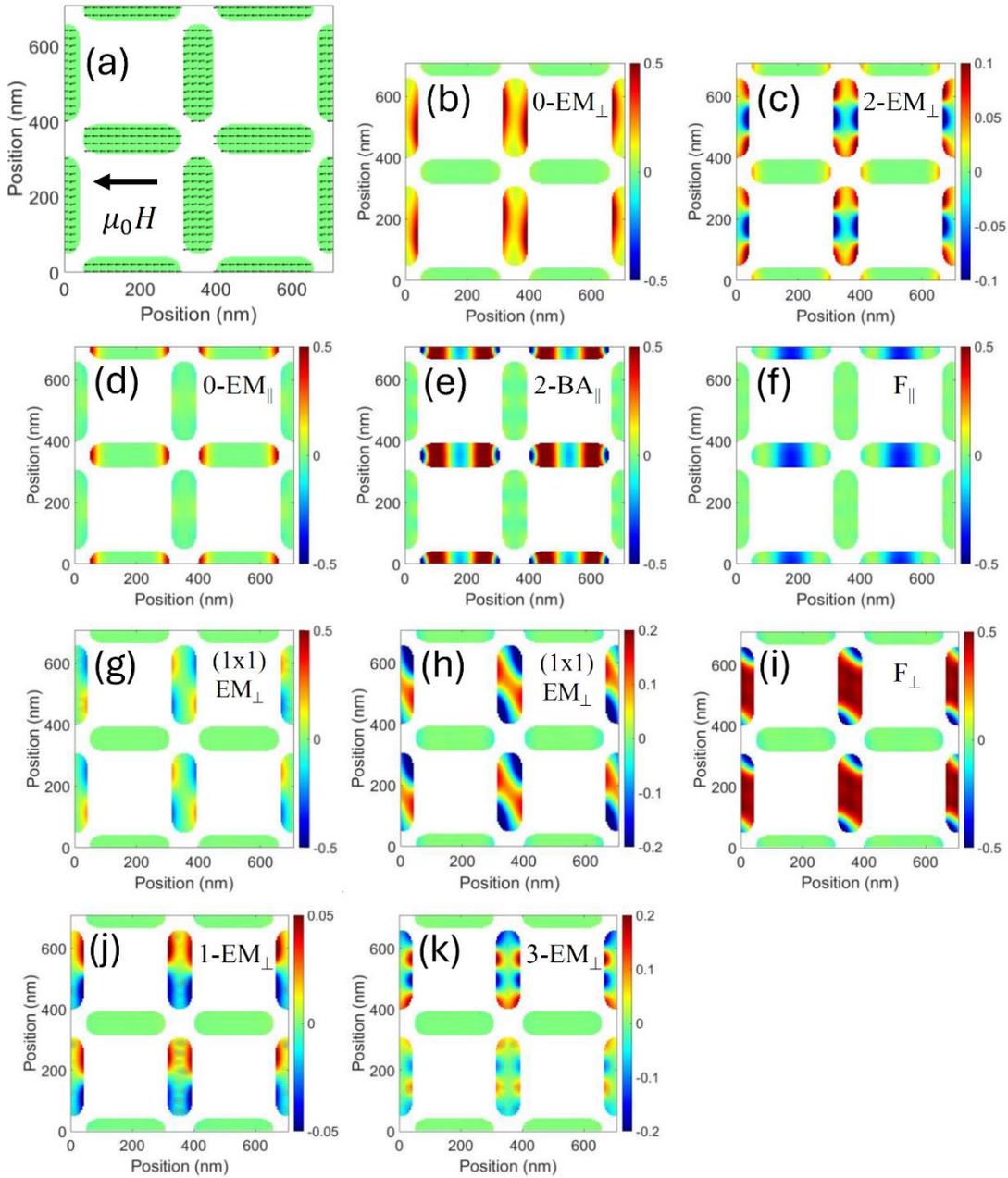

**Figure 6. Magnetization ground state [panel (a), deliberately larger to make the arrows visible] and phase space profile of the main modes of the ASI system (b-k) when $\mu_0 H$=-350 mT. Each**

**profile corresponds to a 2×2 non-primitive unit cell. Panel (b) represents mode 0-EM$_\perp$, at 8.80 GHz, (c) is mode 2-EM$_\perp$, at 20.90 GHz, (d) corresponds to mode 0-EM$_\parallel$, at 21.85 GHz, (e) is mode 2-BA$_\parallel$, at 33.45 GHz, (f) is mode F$_\parallel$, at 34.40 GHz. Note that mode (c) is weakly hybridized with mode (d). In Panel (g-i) we show the evolution of edge mode (1x1)-EM$_\perp$ to F$_\perp$ as the underlying magnetization rotates: (g) 15.65 GHz, at -350 mT; (h) 17.3 GHz, at -250 mT; (i) 19.5 GHz, at -100 mT. Panel (j) and (k) show, correspondingly, the edge modes 1-EM$_\perp$ (15.75 GHz) and 3-EM$_\perp$ (25.5 GHz) calculated at 350 mT, which are usually of negligible intensity, but become BLS active at large incidence angles (see text).**

The frequencies of these three peaks evolve linearly with the applied field, and their frequency separation remains nearly constant within the detectable frequency range: in the experimental BLS measurements, approximately 1.2 GHz between $T_1$ and $T_2$, and 1.5 GHz between $T_1$ and $T_3$. This observation, along with the fact that the measured spectrum of the hybrid structure is not merely a superposition of the spectra from the CoFeB ASI and the NiFe film, is an experimental consequence of both the static and dynamic coupling between ASI and film layers. Regarding static coupling, the ASI nanoelements retain nearly the same magnetization as in the isolated system, owing to their strong shape anisotropy. In contrast, the film layer is significantly influenced by the presence of the ASI layer. Not only does the film acquire the periodicity of the ASI lattice, but also it develops a strong magnetization inhomogeneity inside the primitive cell. This is due to the dipolar interaction, which leads to an overall reduction of all magnon frequencies.

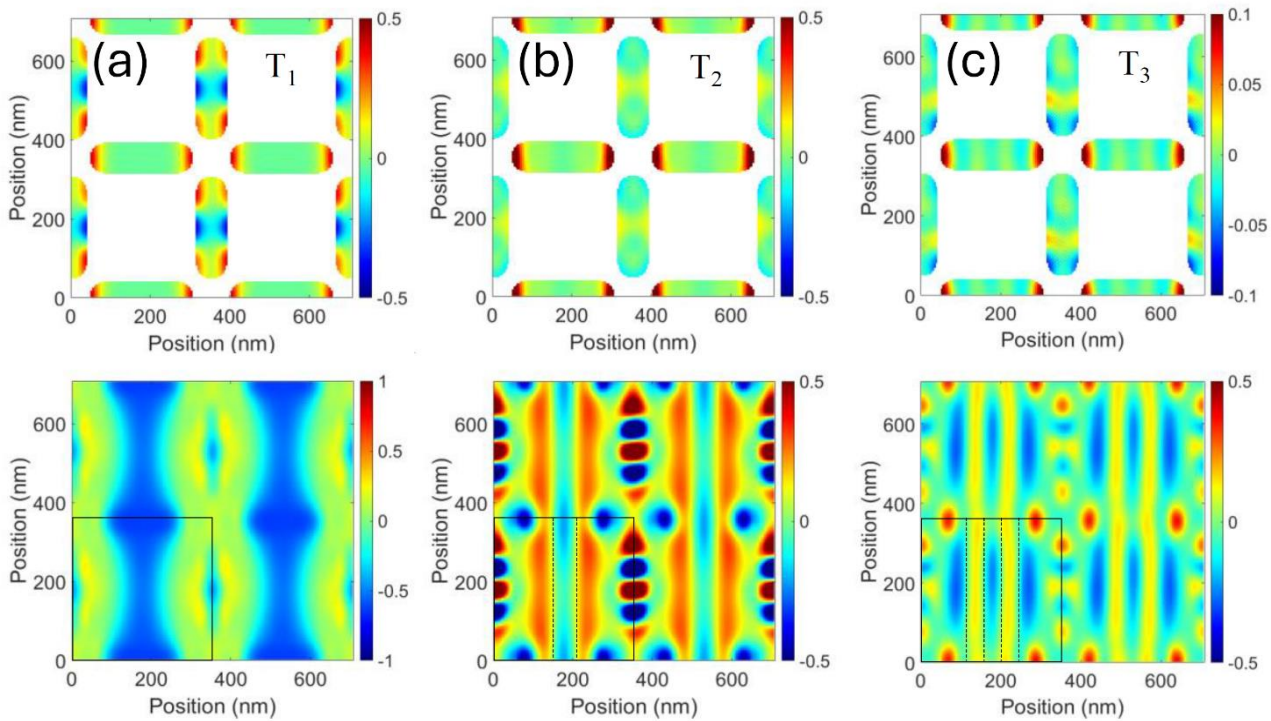

**Figure 7.** Simulated phase space profile of the triplet $T_1$ (a), $T_2$ (b), $T_3$ (c), emerging in the hybrid structure at $\mu_0 H$ =-350 mT. Each profile corresponds to a 2×2 non-primitive unit cells. Top panel: CoFeB ASI layer modes; bottom panel: NiFe film layer modes. Note that the excitation phase profile in the ASI layer is almost the same for all modes $T_1$, $T_2$, or $T_3$, while the film layer phase profiles show an increasing number $m$ of nodes ($m$=0, 2, 4) in $x$-direction selecting $k$=0, $k$=3.57·$10^7$ rad/m, $k$=7.14·$10^7$ rad/m (with $k$ parallel to $\mu_0 H$), respectively, with frequency (a) 20.25 GHz ($T_1$); (b) 21.45 GHz, ($T_2$); (c) 23.55 GHz ($T_3$). The square box on the left-bottom of each panel marks the primitive unit cell, and the dotted lines illustrate the nodal lines.

In terms of dynamic coupling, simulations of our hybrid system show that it occurs exclusively between the ASI island edge modes and the film's propagating backward-volume mode, confined to a narrow frequency window around 22 GHz. More specifically, the two ASI modes 2-EM$_\perp$ and 0-EM$_\parallel$ [in the isolated ASI, at 20.9 GHz and 21.85 GHz, Fig. 6(c,d)] "merge" (i.e., they are excited at the same frequency) in the ASI layer of the hybrid structure and couple with the NiFe film mode (occurring for k=0 at 20.25 GHz in the isolated thin film), resulting in a triplet at (a) 20.25 GHz, (b) 21.45 GHz, (c) 23.55 GHz (Fig. 7). As can be seen from Fig. 7, the phase profile of the ASI layer for these three modes are rather similar (i.e., a superposition of mode 2-EM$_\perp$ and 0-EM$_\parallel$), while the film layer of the same structure exhibits an increasing number of nodal lines (i.e., wavevector), along the direction of the applied field: the nodal lines precisely correspond to the phase distribution observed in the ASI layer. This observation manifests a unique type of magnon-magnon coupling, which can be interpreted as a form of hybridization that persists over a wide range of applied magnetic fields (from 400 mT to 100 mT in our system). This magnon-magnon coupling is enabled by the distinct edge mode characteristics of the horizontal and vertical sublattice sites, their interaction in the presence of the underlayer film, and their coupling with the uniform mode of the NiFe underlayer. Interestingly, this process leads to an excitation of higher wavevectors, which would not be possible to excite without the ASI in the system.

The hybrid mode $T_1$ at 20.25 GHz [Fig. 7, panel (a)] has no full nodal lines in the film layer (bottom panel), while $T_2$ at 21.45 GHz [panel (b)] has (with reference to the primitive unit cell) two nodal lines. Finally, $T_3$ at 23.55 GHz has 4 nodal lines. The number of nodes in the film layer, which in this case is along the direction of the applied field, determines an effective wavelength of infinite, a half and a quarter of the unit cell, respectively, from which we get the effective wavevectors $k$=0,[31] $k$=3.57·$10^7$ rad/m, and $k$=7.14·$10^7$ rad/m. By inserting these wavevector values in the dipole-exchange analytical dispersion of the backward volume SWs (i.e., with wavevector parallel to the applied field),[32] we obtain the expected frequencies for the corresponding uncoupled (uniform) film, namely: 20.32 GHz, 19.96 GHz, and 22.55 GHz, respectively. This analysis enables us to determine the frequency region where the coupling occurs, as well as the frequency spread of the coupled modes. The magnon-magnon interaction can thus be interpreted as a perturbation that mixes the original mode profiles and shifts their frequencies.

Now we pause the discussion to outline the general theory of the dynamic coupling.[13,15] In general, the excitations of two coupled magnetic layers remain independent when external magnetic or geometric parameters, such as applied field or film thickness, are varied. However, under specific conditions,[13,15] a dynamic coupling emerges between the excitations of the two layers, manifesting as an interaction among two or more magnons that remains robust against changes in geometry or magnetic field variations. In such cases, a specific ASI layer mode profile couples to a specific film layer mode profile at the same frequency, so that both are excited simultaneously—in other words, they become "locked." This coupling persists over a specific parameter range, whose width scales with the interaction strength. Consequently, the extent of this range provides an indirect experimental measure of the coupling. The resulting magnon–magnon interaction leads to hybridization of the ASI and film modes, potentially giving rise to distinct acoustic and optical branches. As discussed in more detail in Refs.[13,15], the dynamic coupling is facilitated by mode hybridization and is most likely to occur between mode profiles with compatible symmetry, i.e., the same distribution of nodal lines (same odd/even number of nodal lines) or relative phase. Notably, proximity in frequency of modes is a necessary but not sufficient condition for dynamic coupling to take place.

Let us now return to the present investigation. The additional experimentally observed BLS peaks, specifically peaks (1) and (3) in Fig. 5(b), have so far been left out from the discussion. The comparison with the simulations, based on proximity in frequency and similarity in the slope of the curves, provides insights into their origin. The lowest frequency peak (3) is attributed to mode 0-EM$_\perp$ [Fig. 8(a)]. Despite being classified as an edge mode, it exhibits considerable intensity. As evident from the calculated profile, the amplitude is strongly enhanced not only at the field-aligned edges - where the effective field exhibits deep minima - but also at the island center, indicating hybridization with the fundamental mode. Its field behavior [Fig. 5(b)] closely resembles that observed in the ASI alone, suggesting that it still originates from the ASI layer only (whose magnetization distribution is barely affected by the static coupling with the film). The associated film layer does not contribute to the overall excitation. In fact, by inspection of the corresponding film layer [Fig. 8(a) bottom panel] we recognize only negligible dynamic magnetization. The only effect of the film layer on the ASI is a slight frequency downshift, from 8.8 GHz to 8.6 GHz, due to the *static* dipolar interaction between the layers.

Similarly, peak (1), which appears at the highest measured frequency in Fig. 5(b), exhibits nearly the same absolute values and frequency-field dependence as in the uncoupled ASI [Fig. 5(a)], indicating that it originates solely from the ASI layer. By comparison with the simulations, we attribute peak (1) to the modes shown in Fig. 8(b) and (c), which lie close in frequency and can therefore explain the observed peak broadening. These correspond to the bulk modes 2-BA$_\parallel$ and F$_\parallel$ in the ASI layer. In contrast, the film layer remains essentially unexcited, as evidenced by the negligible dynamic magnetization in its profile. The only contribution of the film is a slight frequency shift due to static (dipolar) coupling: mode (b) is upshifted from 33.45 GHz to 33.6 GHz, while mode (c)—the ASI fundamental—is downshifted from 34.4 GHz to 34.2 GHz.

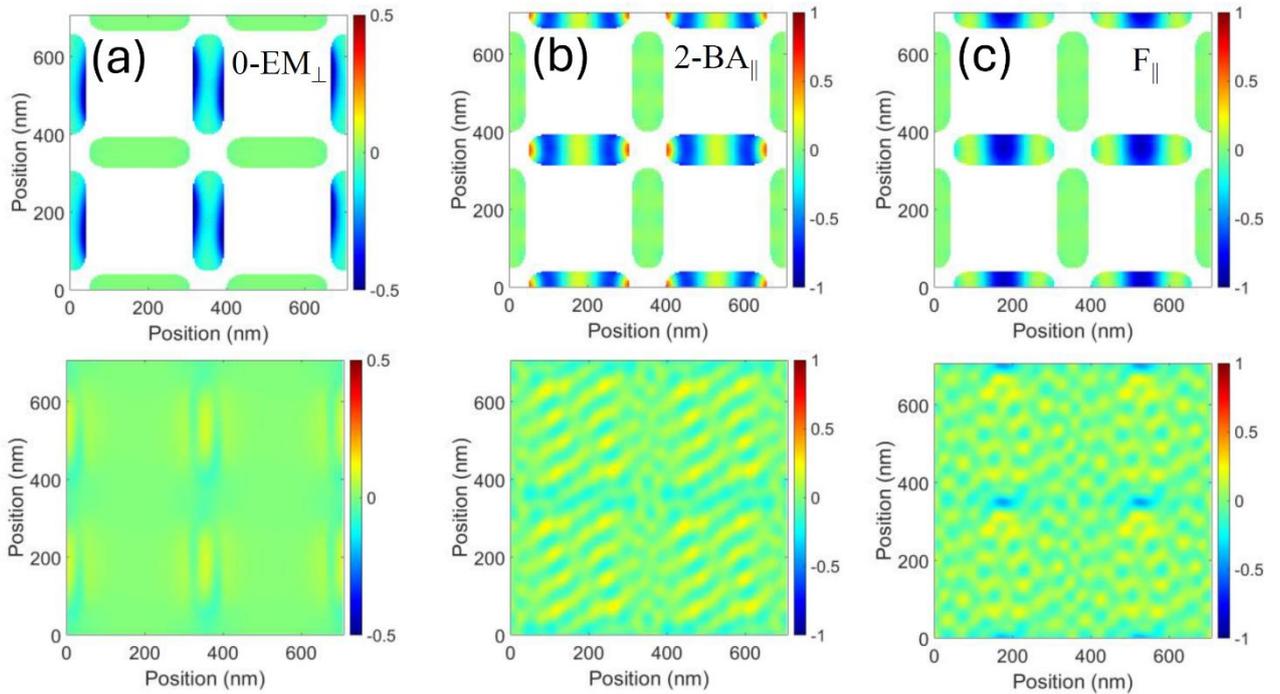

**Figure 8. Phase profile of the top and bottom layers in the hybrid ASI/film structure calculated at an applied magnetic field of 350 mT. Top layers show the phase profiles of ASI modes (a) 0-EM$_\perp$, at 8.60 GHz; (b) 2-BA$_\parallel$, at 33.6 GHz; and (c) F$_\parallel$, at 34.2 GHz. In all three cases, the bottom layer remains unresponsive although subjected to microwave excitation, it exhibits negligible dynamic magnetization, suggesting that ASI and film are dynamically uncoupled at these frequencies.**

SW frequency dispersions (frequency vs $k$) were measured for all samples to investigate the role of interlayer dynamic coupling between the SW modes in the NiFe film and the CoFeB ASI islands. Figure 9 shows a sequence of BLS spectra measured for different wavevector values, comparing the ASI system (Sample #3) with the ASI/film hybrid system (Sample #4).

In the CoFeB ASI system, three prominent peaks, labeled as 1, 2 and 3, are observed in the low-wavevector range in the BLS spectra. Additionally, two more peaks (4 and 5)—located at 14.7 GHz and 24.36 GHz—appear when the wavevector $k$ exceeds $0.73\times10^7$ rad/m. At $k=0$, the lowest-frequency peak, centered around 10.2 GHz, is relatively broad, whereas the peaks at approximately 20 GHz and 35.5 GHz are sharper and more well-defined.

To explain the appearance of additional peaks at higher wavevectors, it is important to consider the wavevector dependence of the BLS signal, which is governed by the spatial distribution of the dynamic magnetization. BLS measurements can be interpreted within the framework of the so-called *Fourier microscopy approach*: at small wavevectors (i.e. $\theta_i$ close to zero), the BLS scattering cross-section is proportional to the squared modulus of the Fourier transform of the out-of-plane component of the dynamic magnetization.[33,34,35,36]

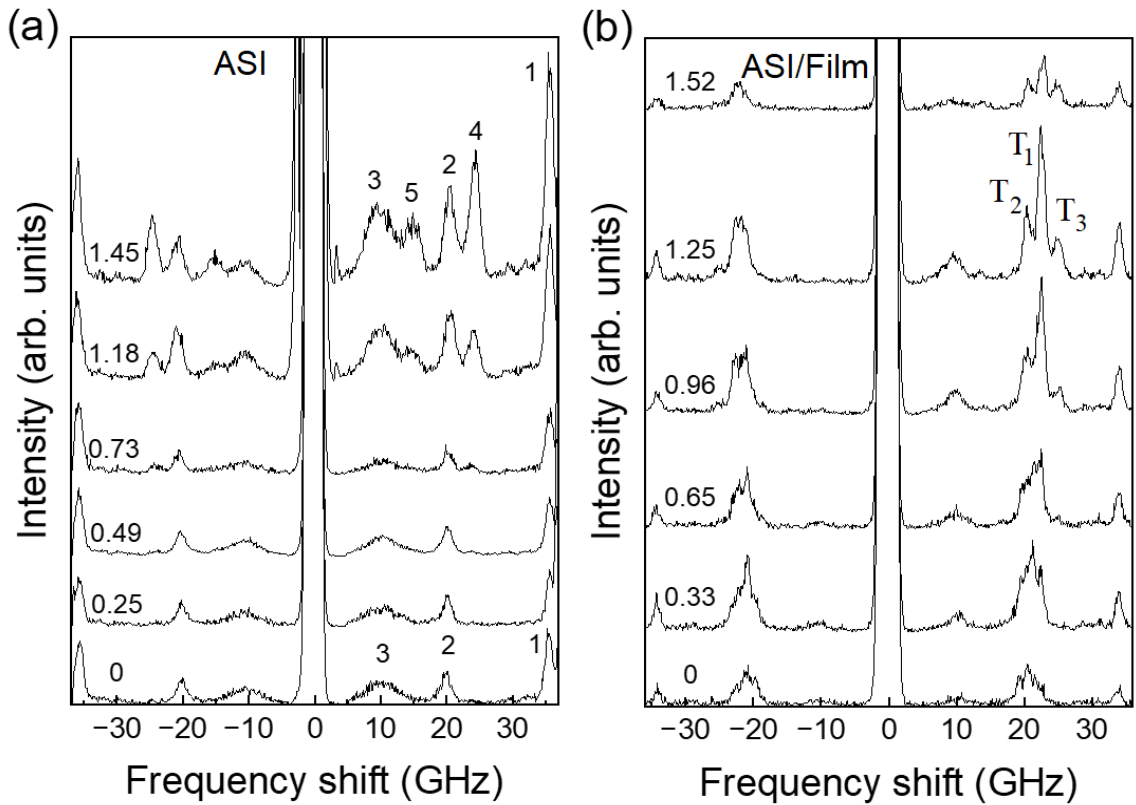

**Figure 9. Comparison between wavevector-resolved BLS spectra of (a) single-layer CoFeB ASI (Sample #3) and (b) CoFeB ASI/NiFe film hybrid structure (Sample #4) at $\mu_0H$ = 350 mT applied along the ASI symmetry direction. The numeric labels of each spectrum (on the left side) indicate the corresponding $k$-values (in units of $10^7$ rad/m). Numeric and alphabetic labels on the right side correspond to the peaks discussed in the text.**

Propagating modes in a periodic medium can be expressed as Bloch waves because of the translational symmetry[37,38]:

$$\partial m_k(r) = \partial \tilde{m}_k(r) e^{ikr} \quad \text{Eq. (2)}$$

where $\partial m_k$ is the actual dynamic magnetization, $\partial \tilde{m}_k(r)$ is the cell function (limited to the primitive unit cell), $r$ is the position across the lattice, and $k$ is the SW wavevector (in our case, perpendicular to the applied field). The BLS technique detects the out-of-plane component of $\partial m_k(r)$, which varies with the wavevector according to Eq. (2). It was shown in Ref. [35] that the BLS amplitude can be written as a product of a form factor (calculated over the illuminated area, and hence involving many primitive unit cells and depending on the primitive lattice vector $\boldsymbol{R}$) and a structure factor, calculated at $\boldsymbol{R} = 0$ limited to a single primitive unit cell. For a sufficiently large illuminated area, the form factor reduces to $\delta(\Delta q - k)$ where:

$$\Delta q = \frac{4\pi}{\lambda} \sin\theta_i$$

is the in-plane transferred wavevector of the probing light beam and $k$ is the magnon wavevector, and $\delta$ is the Dirac delta function. This implies $k = \Delta q$, which corresponds to the matching of the in-plane transferred photon momentum and the magnon momentum. The structure factor, on the other hand, is limited to a single primitive unit cell ($\boldsymbol{R} = 0$), where $\partial m_k(r) \equiv \partial \tilde{m}_k(r)$, and reads:

$$S = \int_{cell} \partial \tilde{m}_k(r) e^{i\Delta q \cdot r} dr \quad \text{Eq. (3)}$$

The final BLS intensity is $I_{BLS} \sim |S|^2$ if $k = \Delta q$.

For modes with an even phase profile $\partial \tilde{m}_k(r)$, the BLS signal is the strongest near the normal incidence (i.e., around $\theta_i = 0°$), and such modes are visible over a wide range of incident angles. In contrast, modes that exhibit an odd phase profile in the wavevector direction, characterized by an effective wavelength $\Lambda$ and wavevector $k = \frac{2\pi}{\Lambda}$, lead to a resonant enhancement of the BLS signal at specific incidence angles. As such, the emergence of high-$k$ peaks in the spectrum can be attributed to any modes that become optically accessible only at larger incidence angles, corresponding to larger transferred wavevectors.

The mode profiles we show in Figs. 6, 7, and 8 correspond to a 2×2 repetition along $x$ and $y$ of the real part of the $z$-component of the SW mode cell function $\partial \tilde{m}_k(r)$, i.e., the SW mode profile $\partial m_k(r)$ at $k = 0$. When $k$ (i.e., $\theta_i$) is increased, the exponential Bloch factor in Eq. (2) introduces a progressive phase shift in $\partial m_k(r)$ which gradually changes the symmetry of the SW profile $\partial m_k(r)$, and hence impacts the BLS cross section and overall signal intensity as discussed above[35,36].

Comparison with the simulations helps to associate peaks 4 and 5, (measured at 14.7 GHz and 24.36 GHz at 350 mT) with edge modes having an odd number of nodes perpendicular to the applied field, specifically mode 1-EM$_\perp$ and 3-EM$_\perp$ [Fig. 6 (j,k), calculated at 350 mT at 17.75 GHz and 25.5 GHz, respectively]. These modes are localized in islands oriented perpendicular to the field, where oscillations exist along the island long axis. If these modes were instead confined to islands parallel to the field, the oscillations would have been confined within the island short axis, resulting in significantly higher frequencies. The ASI modes 1-EM$_\perp$ and 3-EM$_\perp$ exhibit odd cell functions $\partial \widetilde{m}_k(r)$ because those modes have odd nodal lines perpendicular to the field. For an effective spin-wave wavelength $\Lambda$ that equals the lattice constant of the artificial spin-ice lattice $a = 352$ nm, the Bloch wavevector is $k = \frac{2\pi}{a}$: for those odd modes, this results in an even Bloch SW profile $\partial m_k(r)$, because at $r = \frac{a}{2}$, the product of the phase factor $e^{ikr}$ and $\partial \widetilde{m}_k(r)$ is positive [see Eq. (3)], which means the probing light for a particular incidence angle and the magnon are in phase, leading to a resonant enhancement of the signal. However, in the experiment, the BLS cross section appears to become non-vanishing well below this value: in the experiments [Fig. 10(a)], this enhancement begins already across the first Brillouin zone boundary.

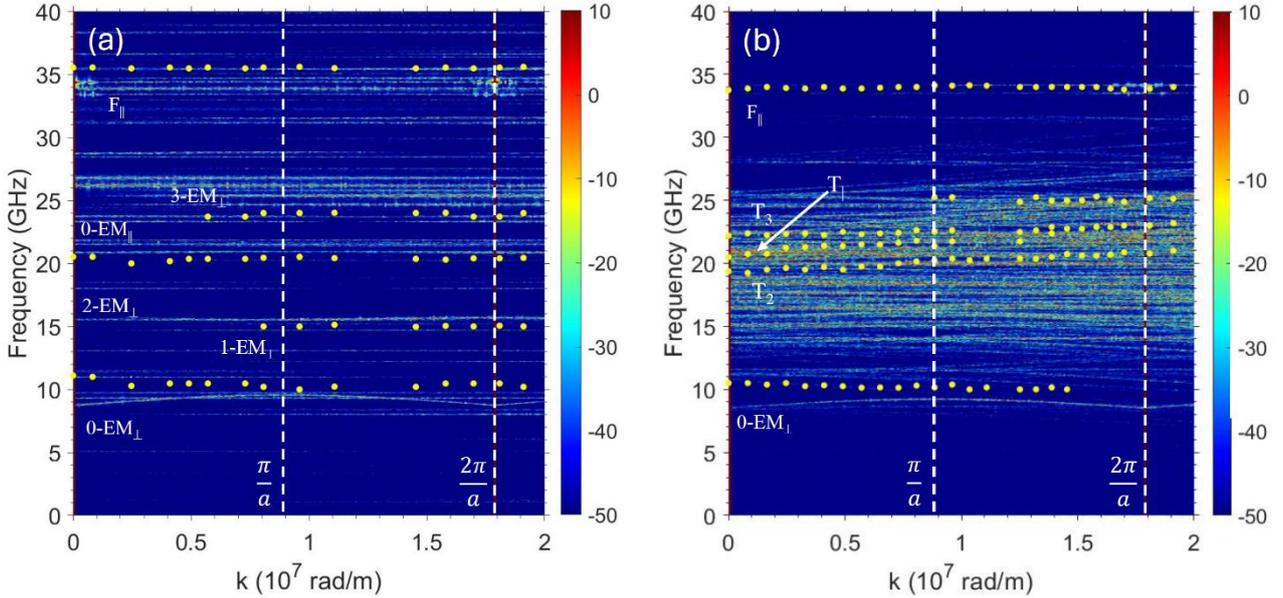

**Figure 10. Comparison between the measured (data points) and simulated (color map) SW frequency dispersion for the ASI [panel (a)] and ASI-Film hybrid [panel (b)], for an applied field $\mu_0 H$ =350 mT. The vertical dashed lines mark the boundaries of the first and second Brillouin zone. A propagating mode (i.e., with non-vanishing group velocity) appears in the hybrid system with frequency around 20 GHz at *k*=0 (indicated by the arrow), called T$_1$ in the text.**

Figure 10 compares the measured and simulated dispersion relation between 0 to $2.0\times10^7$ rad/m corresponding to the 1$^{st}$ and 2$^{nd}$ Brillouin zones in the reciprocal space for the ASI system (Sample #3) and the hybrid sample (Sample #4). For the ASI sample, all modes are dispersionless, as is shown in panel (a); that is, their frequencies remain constant with varying wavevector $k$, indicating that the SW modes do not propagate through the ASI structure due to the rather small inter-element dynamic coupling between the modes originating from the vertical and horizontal islands.

For the hybrid ASI-film structure [Fig. 9 (b)], inspection of the BLS sequence of spectra indicate that the lowest frequency mode (9.7 GHz) and the highest frequency mode (34.0 GHz) exhibit minimal variations across the measured $k$ range, suggesting they originate from stationary modes of the ASI layer stationary modes (in agreement with the other evidences discussed above). The peak at 9.7 GHz is broad and relatively weak in intensity: both features are consistent with the "edge" character of mode 0-EM$_\perp$ in Fig. 8(a) found in the simulations. The peak at 34.0 GHz is sharp and well-defined, consistent with the "bulk" character of the mode F$_\parallel$ shown in Fig. 8(c).

Conversely, the mode triplet observed between about 18.7 and 22.9 GHz, exhibits a significant frequency evolution as $k$ increases, with the frequency separation between peaks increasing with $k$. This trend is particularly noticeable on the anti-Stokes side of the spectra, as can be inferred from the spectra presented in Fig. 9.

Comparison with simulations helps to associate the mode with the largest dispersion slope to the T$_1$ hybrid mode shown in Fig. 7(a), while the other two, less dispersive modes are identified as T$_2$ and T$_3$ modes [Figs. 7(b) and (c)], respectively. This interpretation is supported by the fact that a higher number of nodes in the film layer leads to reduced dynamic stray field and, consequently, narrower bandwidths. The same argument justifies the largest intensity of T$_1$ BLS peak, mentioned previously. In the simulations, the two modes T$_2$ and T$_3$ appear at higher frequencies than T$_1$, while in the experimental data, based on their dispersion slopes, they appear above and below the frequency of the T$_1$ mode.

## CONCLUSIONS

We report the hybridization of a backward spin-wave mode in a ferromagnetic film with edge modes of an artificial spin ice system, resulting from the dynamic coupling between layers made of different magnetic materials. Such hybridization manifests as a mode triplet in the BLS spectra. We interpret this triplet as a distinctive magnon-magnon interaction, facilitated by the large magnetic contrast between NiFe and CoFeB, i.e. difference in the $M_s$ values. As revealed by an extensive analysis of the frequency-field dependencies, this interaction is robust within a wide range of applied magnetic fields. In the dispersion relationships, the triplet of modes exhibits a propagating character, particularly the mode with the highest intensity in the spectra, which is associated with a symmetric dynamic profile and consequently the largest dynamic stray fields. These findings are supported by both experimental measurements and micromagnetic simulations, offering a consistent and comprehensive picture. Together, the results establish a framework in which the careful design of geometry and material composition serves as an additional degree of freedom for controlling signal transmission and manipulation at the nanoscale. Furthermore, the studied system reveals a distinct form of magnon-magnon coupling enabled by vertical nanomagnonic design benefitting the research field of hybrid magnonics.

## METHODS

**Fabrication.** The samples were fabricated by depositing a $Ni_{81}Fe_{19}$ (Permalloy) thin film onto a thermally oxidized silicon substrate using electron beam (e-beam) evaporation. To introduce a non-magnetic spacer layer, we deposit aluminum oxide ($Al_2O_3$) using the same e-beam evaporation technique. The ASI geometry is defined through electron beam lithography (EBL), followed by the deposition of $Co_{40}Fe_{40}B_{20}$ (CoFeB) using e-beam evaporation. A subsequent lift-off process ensures precise patterning of the ASI structure. Further details on the fabrication procedure are provided in the Supplemental Material of Ref. 13.

**Measurements.** Magneto-optical Kerr effect (MOKE) magnetometry, which is based on the change in polarization rotation and ellipticity of light upon reflection from a magnetic medium, has proven to be a powerful technique for measuring magnetization curves in magnetic thin films and multilayers.[39] In our experiments, hysteresis loops were measured at room temperature in the longitudinal configuration, i.e., with an applied magnetic field parallel to the sample surface and along the symmetry direction of the ASI. The magnetic field $\mu_0 H$ was swept in the range from -250 to +250 mT in steps of 4 mT. Our MOKE setup consists of a laser source, a photoelastic modulator operating at 50 kHz, and a lock-in amplifier for precise signal detection and noise reduction.

BLS spectra were recorded at room temperature in a backscattering configuration.[40] A monochromatic, p-polarized laser beam with a wavelength of λ=532 nm and a power of approximately 200 mW was focused onto the sample surface at different incidence angles relative to the sample normal. The backscattered s-polarized light was then analyzed using a (3+3) tandem Fabry–Pérot interferometer,[41] enabling high-resolution spectral measurements of the SW modes. A dc magnetic field was applied parallel the sample plane and perpendicular to the incidence plane of light, which defines the direction of the SW wavevector $k$. The sample was mounted on a goniometer, enabling rotation around the field direction to vary the incidence angle of light ($\theta_i$) from 0° to 70°. Due to the conservation of in-plane momentum in the scattering process, the SW wavevector $k$ is determined by the light incidence angle $\theta_i$ and light wavevector $\frac{2\pi}{\lambda}$, following the relation:

$$k = \frac{4\pi}{\lambda} \sin\theta_i \qquad \text{Eq. (3)}$$

This simple relation establishes a direct correlation between the experimental geometry ($\theta_i$) and the magnon wavevector ($k$), allowing for precise control over the probed SW entering in the scattering process.

**Micromagnetic simulations**

Micromagnetic simulations of the magnetization dynamics in the ASI structures (Sample #3 and #4), micromagnetic simulations were performed with the GPU-accelerated program Mumax3.[42] The micromagnetic elemental cell was set to 4×4×5 nm$^3$, and each island was represented by an oval shape with 64×20 cells, i.e., 256×80 nm$^2$. For the simulations of the frequency/field curves and the mode profiles, a non-primitive 2×2 unit cell was adopted, to account for possible different magnetic charges (magnetization directions) at the ASI vertices,[30] with an area of 704×704 nm$^2$. We excited the system by a sinc pulse, with amplitude 1 mT and cut-off frequency 50 GHz. The simulation duration time was 20 ns, giving a frequency resolution 0.05 GHz, while the sampling timestep was 0.01 ns, giving the maximum (Nyquist) frequency of 50 GHz. The sinc excitation is applied to the relaxed, equilibrium magnetization at each applied field value, from -400 mT to + 400 mT, in field steps of 2 mT. A slight tilt of 1 degree was applied between applied field direction and one of the primary ASI axes, to avoid any computational artifacts and account for unavoidable experimental error in field alignment.

For the frequency-wavevector dispersion at $\mu_0 H$=350 mT, a primitive unit cell (352×352 nm$^2$) could be used, since the large field allowed only two possible magnetizations, corresponding to the two orientations of the islands. In this case, to allow long wavelength phase variations, we used a supercell comprising of 200 of the primitive cells, resulting in a wavevector resolution of ~0.0089×10$^7$ rad/m. The simulation time was 100 ns, corresponding to a frequency resolution of 0.01 GHz, while the sampling timestep was 12.5 ps, corresponding to a maximum frequency of 40 GHz.

The simulated spectra show the square modulus of the out-of-plane component of the Fourier coefficients (in log scale), averaged between film and ASI layers in the hybrid system, while for the mode profiles (shown in Fig. 6, 7 and 8) we plot the real part of the out-of-plane component of the Fourier coefficients (space-resolved phase amplitude),[32,43] limited to the 2×2 non-primitive unit cell.

## AUTHOR INFORMATION
**Corresponding Author**
*E-mail: montoncello@fe.infn.it
**ORCID**
Federico Montoncello: 0000-0002-4811-3675

**Author Contributions**
R. S., M. T. K., and Y. J. prepared the samples. M. B. J. planned the project. G. G. characterized the samples and performed MOKE and BLS measurements. F. M. performed the micromagnetic simulations and the data analysis in discussion. G.G., M. B. J. And F. M. wrote the manuscript in consultation with all authors.

**Notes**
The authors declare no competing financial interest.

## ACKNOWLEDGMENTS
FM acknowledges support by the Department of Physics and Earth Sciences-University of Ferrara Grant Bando FIRD 2024, as well as the CINECA award under the ISCRA initiative, for the availability of high-performance computing resources and support. GG acknowledges funding by the European Union–NextGenerationEU, Mission 4, Component 1, under the Italian Ministry of University and Research (MUR) National Innovation Ecosystem grant ECS00000041-VITALITY-CUP B43C22000470005 and by 'PNRR—M4C2, investimento 1.1—'Fondo PRIN 2022'—TEEPHANY–ThreEE-dimensional Processing TecHnique of mAgNetic crYstals for magnonics and nanomagnetism ID 2022P4485M- CUP D53D23001400001. Work at the University of Delaware including sample fabrication and data analysis was supported by the U.S. Department of Energy, Office of Science, Office of Basic Energy Sciences, under Award No. DE-SC-0024346. MBJ acknowledges the JSPS Invitational Fellowship for Researcher in Japan.

# REFERENCES

(1) Drisko, J.; Marsh, T.; Cumings J. Topological frustration of artificial spin ice. *Nat. Commun.* **2017**, *8*, 14009.

(2) Sklenar, J.; Lao, Y.; Albrecht, A.; Watts, J. D.; Nisoli, C.; Chern G.-W.; Schiffer, P. Field-induced phase coexistence in an artificial spin ice. *Nat. Phys.* **2019**, *15*, 191.

(3) Bhat, V. S.; Heimbach, F.; Stasinopoulos, I.; Grundler, D. Magnetization dynamics of topological defects and the spin solid in a kagome artificial spin ice. *Phys. Rev. B* **2016**, 93, 140401(R).

(4) Skjærvø, S. H.; Marrows, C. H.; Stamps, R. L.; Heyderman, L. J. Advances in artificial spin ice. *Nat. Rev. Phys.* **2019**, *439*, 1.

(5) Iacocca, E.; Gliga, S.; Stamps, R. L.; Heinonen, O. Reconfigurable wave band structure of an artificial square ice. Phys. Rev. B **2016**, *93*, 134420.

(6) Gliga, S.; Iacocca, E.; Heinonen, O. G. Dynamics of reconfigurable artificial spin ice: Toward magnonic functional materials. *APL Mater*. **2020**, *8*, 040911.

(7) Lendinez, S.; Jungfleisch, M. B. Magnetization dynamics in artificial spin ice. *J. Phys.: Condens. Matter* **2020**, *32*, 013001.

(8) Kaffash, M. T.; Lendinez, S.; Jungfleisch, M. B. Nanomagnonics with artificial spin ice. *Physics Letters A* **2021**, *402*, 127364.

(9) Sultana R.; et al., Ice sculpting: An artificial spin ice Tutorial on controlling microstate and geometry for magnonics and neuromorphic computing. *J. Appl. Phys.* **2025**, *138*, 061101.

(10) Lendinez, S.; Kaffash, M. T.; Heinonen, O. G.; Gliga, S.; Iacocca, E.; Jungfleisch, M. B. Nonlinear multi-magnon scattering in artificial spin ice. Nat. Comm. **2023**, *14*, 3419.

(11) Dion, T.; Stenning, K. D.; Vanstone, A.; Holder, H. H.; Sultana, R.; Alatteili, G.; Martinez, V.; Kaffash, M. T.; Kimura, T.; Oulton, R. F.; Branford, W. R.; Kurebayashi, H.; Iacocca, E.; Jungfleisch, M. B. Ultrastrong magnon-magnon coupling and chiral spin-texture control in a dipolar 3D multilayered artificial spin-vortex ice. *Nat. Comm.* **2024**, *15*, 4077.

(12) Bhat, V. S.; Jungfleisch, M. B. Magnon signatures of multidimensional reconfigurations in multilayer square artificial spin ices. *Appl. Phys. Lett.* **2025**, *126*, 022406.

(13) Negrello, R.; Montoncello, F.; Kaffash, M. T.; Jungfleisch, M. B.; Gubbiotti, G. Dynamic coupling and spin-wave dispersions in a magnetic hybrid system made of an artificial spin-ice structure and an extended NiFe underlayer. *APL Materials*, **2022**, *10*, 091115.

(14) Flebus B. et al. The 2024 magnonics roadmap. *J. Phys.: Condens. Matter* **2024**, *36*, 363501.

(15) Montoncello, F.; Kaffash, M. T.; Carfagno, H.; Doty, M. F.; Gubbiotti; G.; Jungfleisch, M. B. A Brillouin light scattering study of the spin-wave magnetic field dependence in a magnetic hybrid system made of an artificial spin-ice structure and a film underlayer. *J. Appl. Phys.* **2023**, *133*, 083901.

(16) Barker, O. J.; Mohammadi-Motlagh, A.; Wright, A. J. ; Batty, R. ; Finch, H.; Vezzoli, A.; Keatley, P. S. ; O'Brien, L. *Appl. Phys. Lett.* **2024**, *124*, 112411.

(17) Vogel, M., *et al.* Optically reconfigurable magnetic materials. *Nature Physics* **2015**, *11*, 487.

(18) Riddiford, L. J.; Brock, J. A.; Murawska, K.; Hrabec, A.; Heyderman, L. J. Grayscale control of local magnetic properties with direct-write laser annealing. arXiv:2401.09314v1.

(19) Gubbiotti, G. Three-Dimensional Magnonics: Layered, Micro- and Nanostructures (Jenny Stanford Publishing, Singapore, 2019) https://www.jennystanford.com/9789814800730/three-dimensional-magnonics/.

(20) Gubbiotti, G.; Zhou, X.; Haghshenasfard, Z.; Cottam, M. G.; Adeyeye, A.O. Reprogrammable magnonic band structure of layered Permalloy/Cu/Permalloy nanowires. *Phys. Rev. B* **2018**, *97*, 134428.

(21) Gubbiotti, G.; Zhou, X.; Haghshenasfard, Z.; Cottam, M. G.; Adeyeye, A.O.; Kostylev, M. Interplay between intra- and inter-nanowires dynamic dipolar interactions in the spin wave band structure of Py/Cu/Py nanowires. *Sci. Rep.* **2019**, *9*, 4617.

(22) Gubbiotti, G.; Sadovnikov, A.; Beginin, E.; Nikitov, S.; Wan, D.; Gupta, A.; Kundu, S.; Talmelli, G.; Carpenter, R.; Asselberghs, I.; Radu, I. P.; Adelmann, C.; Ciubotaru, F. Magnonic Band Structure in Vertical Meander-Shaped $Co_{40}Fe_{40}B_{20}$ Thin Films. *Phys. Rev. Appl.* **2021**, *15*, 014061.

(23) Graczyk, P.; Krawczyk, M.; Dhuey, S.; Yang, W.-G.; Schmidt, H.; Gubbiotti, G. Magnonic band gap and mode hybridization in continuous permalloy films induced by vertical dynamic coupling with an array of permalloy ellipses. *Phys. Rev. B* **2018**, *98*, 174420.

(24) Micaletti, P.; Roxburgh, A.; Iacocca, E.; Marzolla, M.; Montoncello, F. Magnonic analog of a metal-to-insulator transition in a multiferroic heterostructure. *J. Appl. Phys.* **2025**, *137*, 153906.

(25) Stancil D. D.; Prabhakar, A. Spin Waves Theory and Applications, 1st ed. New York: Springer-Verlag, 2008.

(26) Damon, R. W.; Eshbach, J. R.; Magnetostatic modes of a ferromangetic slab. *J. Phys. Chem. Solids* 1961, *19*, 308.

(27) Bang, W.; Montoncello, F.; Jungfleisch, M. B.; Hoffmann, A.; Giovannini, L.; Ketterson, J. B. Angular-dependent spin dynamics of a triad of permalloy macrospins. *Phys. Rev. B* **2019**, *99*, 014415.

(28) Bang, W.; Montoncello, F.; Kaffash, M. T.; Hoffmann, A.; Ketterson, J. B.; Jungfleisch M. B. Ferromagnetic resonance spectra of permalloy nano-ellipses as building blocks for complex magnonic lattices. *J. Appl. Phys.* **2019**, *126*, 203902.

(29) Goll, D.; Schutz, G.; Kronmuller, H. Critical thickness for high-remanent single-domain configurations in square ferromagnetic thin platelets. Phys. Rev. B **2023**, *67*, 094414.

(30) Micaletti, P.; Montoncello, F. Dynamic footprints of the specific artificial spin ice microstate on its spin waves. *Magnetochemistry* **2023**, *9*, 2312.

(31) We extracted the values from the space Fourier transform of the film layer magnon maps of Fig. 7(b,c), in correspondence of the largest peak of each Fourier spectrum. The wavevector values correspond exactly to wavelength a half and a quarter of the primitive cell size (which is 352 nm).

(32) Venkat, G.; et al., Proposal for a Standard Micromagnetic Problem: Spin Wave Dispersion in a Magnonic Waveguide. *IEEE Transactions on Magnetics* **2013**, *49*, 524.

(33) Demokritov, S. O. *J. Phys.: Condens. Matter* **2003**, *15,* S2575.

(34) Demokritov, S. O.; Hillebrands, B.; Slavin, A. N. Brillouin light scattering studies of confined spin waves: linear and nonlinear confinement. *Physics Reports* **2001**, *348*, 441.

(35) Zivieri, R.; Montoncello, F.; Giovannini, L.; Nizzoli, F.; Tacchi, S.; Madami, M.; Gubbiotti, G.; Carlotti, G.; Adeyeye, A. O. Collective spin modes in chains of dipolarly interacting rectangular magnetic dots. *Phys. Rev. B* **2011**, *83*, 054431.

(36) Gubbiotti, G.; Carlotti, G.; Okuno, T.; Grimsditch, M.; Giovannini, L.; Montoncello, F.; Nizzoli, F. Spin dynamics in thin nanometric elliptical Permalloy dots: A Brillouin light scattering investigation as a function of dot eccentricity. Phys. Rev. B **2005**, *72*, 184419.

(37) Kittel, C. Introduction to solid state physics (8th ed.). (2005) New York: Wiley. p. 680. ISBN 978-0-471-68057-4. OCLC 787838554.

(38) Montoncello, F.; Tacchi, S.; Giovannini, L.; Madami, M.; Gubbiotti, G.; Carlotti, G.; Sirotkin, E.; Ahmad, E.; Ogrin, F. Y.; Kruglyak, V. V. Asymmetry of spin wave dispersions in a hexagonal magnonic crystal. *Appl. Phys. Lett.* **2013**, *102*, 202411.

(39) Qiu, Z. Q.; Bader, S. D. Surface Magneto-optic Kerr effect (SMOKE). *J. Magn. Magn. Mater.* **1999**, *200*, 664.

(40) Carlotti, G.; Gubbiotti, G. Magnetic properties of layered nanostructures studied by means of Brillouin light scattering and the surface magneto-optical Kerr effect. *J. Phys.: Condens. Matter* **2002**, *14*, 8199.

(41) Sandercock, J. R. in *Light Scattering in Solids III*, ed. by M. Cardona, G. Guntherodt (Springer Series in Topics.

(42) Vansteenkiste, A.; Leliaert, J.; Dvornik,.; Helsen, M.; Garcia-Sanchez, F.; Van Waeyenberge, The design and verification of MuMax3. *AIP Adv.* **2014**, *4*, 107133.

(43) Van de Wiele, B.; Montoncello, F. A continuous excitation approach to determine time-dependent dispersion diagrams in 2D magnonic crystals. *J. of Physics D: Applied Physics* **2014**, 47, 315002.